\documentclass[aps,floatfix,reprint,superscriptaddress]{revtex4-2}
\usepackage{amsmath,amssymb,graphicx,mathtools}
\usepackage{float}

\usepackage[T1]{fontenc}

\usepackage{color}
\usepackage{colortbl}
\usepackage{nicefrac}
\usepackage[colorlinks]{hyperref}
\usepackage{tabularx}

\usepackage{calligra}

\usepackage{tikz-cd}

\usepackage{multirow}

\usepackage{diagbox}

\usepackage{scalerel}
\usepackage{stackengine,wasysym}

\makeatletter
\DeclareRobustCommand{\cev}[1]{%
  {\mathpalette\do@cev{#1}}%
}
\newcommand{\do@cev}[2]{%
  \vbox{\offinterlineskip
    \sbox\z@{$\m@th#1 x$}%
    \ialign{##\cr
      \hidewidth\reflectbox{$\m@th#1\vec{}\mkern4mu$}\hidewidth\cr
      \noalign{\kern-\ht\z@}
      $\m@th#1#2$\cr
    }%
  }%
}
\makeatother

\newcounter{theoremcounter}
\stepcounter{theoremcounter}

\newcommand{\bs}{\boldsymbol}

\definecolor{blue(ryb)}{rgb}{0.01, 0.28, 1.0}
\definecolor{scarlet}{rgb}{1.0, 0.13, 0.0}

\makeatletter
\newcommand{\vast}{\bBigg@{4}}
\newcommand{\Vast}{\bBigg@{5}}
\makeatother

\begin{document}

\title{AdS$_3$ Vacuum State from Four Minkowski Vacuum States}
\author{Lucas Kocia Kovalsky}
\affiliation{Quantum Algorithms and Applications Collaboratory, Sandia National Laboratories, Livermore, California 94550, U.S.A.}
\date{\today }
\begin{abstract}
  We show that a tensor product of four specific $1{+}2$ Minkowski vacuum states is a self-consistent vacuum state for an infinite set of three-dimensional anti-de Sitter (AdS$_3$) spacetimes if their parity and time-reversal symmetry are broken in a particular way. The infinite set consists of pairs of all AdS$_3$ with non-zero unique scalar curvatures. %
\end{abstract}
\maketitle

While a vacuum state can be defined for quantum fields in Minkowski spacetimes, %
consistent definitions have only been found for curved spacetimes that are globally hyperbolic. Such manifolds possess a Cauchy surface with a domain of dependence that covers them fully. Without a Cauchy surface, the non-linear Einstein equations of general relativity lack a complete set of initial value conditions~\cite{Carroll19}.

There have been many attempts to develop a more general class of vacuum states. These include constructions for static spacetimes with respect to measurements along proper time~\cite{Ishibashi04,Carroll19} and the S-J vacuum state~\cite{Johnston09,Sorkin11,Afshordi12}, among others~\cite{Kachru03,Narayan10,Ooguri16}. There have also been efforts to algebraically define an equivalent class of states, such as the Hadamard state in the GNS construction~\cite{Wald94}. However, so far, results generally lack all the features of vacuum states in globally hyperbolic spacetimes~\cite{Allen85,Wald10,Fewster12}. This challenge is one of the main outstanding obstacles to formulating a consistent theory of quantum gravity. Here, we will show that perhaps the key to generalization lies in expressing an infinite set of curved subspaces instead as a direct sum of a finite number of flat projected spaces. Furthermore, we will show that AdS's boundary, which is of central importance to the AdS/CFT correspondence but also prevents global hyperbolicity, can sometimes be tamed by breaking the parity and time-reversal symmetry of AdS in a particular way to produce a globally hyperbolic spacetime, while still preserving the key features of the AdS/CFT correspondence.

We will accomplish this by developing a quantum field theory inspired by the curved spacetime emergently defined by a non-commutative algebra~\cite{Connes96,Aschieri06,Connes19,Singh20}. In particular, our construction will respect the Cartan decomposition of a recent formalism, which produced a consistent stress-energy tensor for some non-globally hyperbolic spacetimes using the octonion algebra~\cite{Kovalsky23}. We will consider the simplest case: three-dimensional anti-de Sitter (AdS$_3$) spacetime.

\begin{figure}[t]
\includegraphics[]{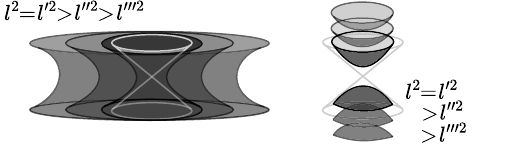}
\caption{A sketch of the AdS$_2$ embedded spacetimes in a flat \(1{+}2\) (left) and \(2{+}1\) (right) embedding space. The oppositely signed signatures restrict foliation to the outside or inside of the $1{+}2$ embedding's lightcone, respectively. In AdS$_3$ the left and right foliations are non-overlapping inverted copies of each other.
}
\label{fig:embeddingembeddedspace}
\end{figure}

Consider a \(2{+}2\) flat spacetime with metric \(ds^2 = -dt^2 + dx^2 - dt'^2 + d x'^2\) and embed AdS$_3$ spacetime by restricting to the hyperboloid \(-t^2 + x^2 - t'^2 + x'^2 = -l^2\) for \(l^2{>}0\). Equivalently, one can consider flipping the signature to the metric \(ds^2 = dt^2 - dx^2 + dt'^2 - dx'^2\) and restricting to the hyperboloid \(-t^2 + x^2 - t'^2 + x'^2 = l^2\) (see Fig.~\ref{fig:embeddingembeddedspace}). Selecting the first signature henceforth, the six isometries of the embedding flat spacetime (excluding parity $\mathcal P$ and time-reversal $\mathcal T$ symmetry) are the usual rotations and boosts generated by \(K^{01}\), \(K^{23}\), \(K^{12}\), \(K^{03}\), \(J^{02}\), and \(J^{13}\), where \(K^{01} = t \partial_x + x \partial_t\), \(K^{23} = t' \partial_{x'} + x' \partial_{t'}\), \(K^{12} = x \partial_{t'} + t' \partial_x\), \(K^{03} = t \partial_{x'} + x' \partial_t\), \(J^{02} = t \partial_{t'} - t' \partial_t\), and \(J^{13} = x \partial_{x'} - x' \partial_{x}\). The embedded AdS$_3$ spacetime inherits these isometries. As a result, this set forms the isometry group \(\text{SL}(2,\mathbb R) \times \text{SL}(2,\mathbb R)\) of the double cover of the embedded AdS$_3$ manifolds, restricted to one side or the other of the $2{+}2$ embedding spacetime's lightcone. Each commuting \(\text{SL}(2,\mathbb R)\) is generated by even/odd linear combinations of the former isometries: \(\{K^{01} + K^{23},\, K^{03} + K^{12},\, J^{02} + J^{13}\}\) generate one and \(\{K^{01} - K^{23},\, K^{03} - K^{12},\, J^{02} - J^{13}\}\) generate the other~\cite{Behrndt99}.

Similar embeddings of AdS$_{n+1}$ can be made in $2{+}n$ flat spacetimes. Generally, the generators of embedding spaces form an overcomplete basis for their embedded manifolds and cannot consistently generate them with their integral curves~\footnote{i.e. they do not form a Lie algebra and thus do not generate integral curves by Frobenius' theorem}. However, for $n=2$, the generators of a flat $2{+}n$ embedding space do produce consistent integral manifolds of AdS$_{n+1}$. This can be seen more easily by explicitly treating the $2{+}2$ embedding space as a homogeneous space where transformations look the same up to $\text{SL}(2,\mathbb R)$, thereby producing a quotient space for the embedded manifolds: \(\frac{\text{SL}(2,\mathbb R) \times \text{SL}(2, \mathbb R)}{\text{SL}(2, \mathbb R)} \cong \text{SL}(2, \mathbb R)\). The new space has isometry generators corresponding to the former commuting sets of generators identified with each other---\(K^{01} \pm K^{23}\), \(K^{03} \pm K^{12}\), and \(J^{02} \pm J^{13}\)---and becomes a homogeneous symmetric space~\cite{Helgason79}.

These three generators form a complete basis for \emph{both} the $2{+}2$ embedding space and each three-dimensional AdS$_3$ manifold (see Appendix~\ref{app:AdS3Killingvecs} for a derivation involving this space's Cartan decomposition). In other words, each hyperboloid \(-t^2 + x^2 - t'^2 + x'^2 = -l^2\) precisely coincides with a manifold produced by the integral curves of the quotient \(2{+}2\) spacetime's isometries: \(\{K^{01} + K^{23}\), \(K^{03} + K^{12}\), \(J^{02} + J^{13}\}\) produce clockwise integral curves while \(\{K^{01} - K^{23}\), \(K^{03} - K^{12}\), \(J^{02} - J^{13}\}\) produce counter-clockwise integral curves. \(l^2{>}0\) indexes all such manifolds. Including both signs of the embedding metric results in double covers of two non-overlapping copies (see Fig.~\ref{fig:embeddingembeddedspace}).

The scalar curvature of AdS$_3$ spacetime is \(R = -6/l^2 < 0\). \(l^2\) is often set to \(1\) to represent the ``open'' universe equivalence class since any \(l^2{>}0\) produces the same rescaled physics~\cite{Wald10}. This scale invariance along with the accidental symmetry of the embedding space's signature under inversion is responsible for the completeness of the isometry basis in AdS$_{n+1}$ for $n=2$ without $\mathcal P$ and $\mathcal T$. %

The double cover of an AdS$_3$ embedded spacetime on either side of the $2{+}2$ lightcone is not globally hyperbolic. This is due to the fact that its spatial infinity boundary is a timelike hypersurface. New information can always enter at this open boundary, thereby denying the sufficiency of any initial value conditions~\cite{Carroll19}. Note that while we have only been considering the continuous generators that produce $\text{SL}(2,\mathbb R)$, by considering its action restricted to hyperboloids parametrized by $l^2$, we have still implicitly been including $\mathcal P$- and $\mathcal T$-symmetry. This suggests that perhaps by taking advantage of the freedom in how $\mathcal P$- and $\mathcal T$-symmetry can be removed, we may be able to produce a spacetime that is globally hyperbolic while still preserving its continuous symmetries. %

In particular, in the following, we will see that global hyperbolicity can be satisfied if the $2{+}2$ embedding space is built up with broken $\mathcal P$- and $\mathcal T$-symmetry by using a finite number of flat projections instead of an infinite number of curved subspaces defined by the hyperbolic restriction which preserve $\mathcal P$ and $\mathcal T$-symmetry. %

The complete isometry generator basis' decomposition into pairs of generators can be usefully combined with the ``intensive'' property of homogeneous spaces: projections of homogeneous symmetric spaces are also homogeneous if their kernel is homogeneous. Such homogeneous projections can be defined for the original \(2{+}2\) embedding space by evenly splitting the pairs of generators of its quotient space into the following four sets: \(\{K^{01}\), \(K^{03}\), \(J^{13}\}\), \(\{K^{01}\), \(K^{12}\), \(J^{02}\}\), \(\{K^{23}\), \(K^{03}\), \(J^{02}\}\), and \(\{K^{23}\), \(K^{12}\), \(J^{13}\}\). Each set consists of a sub-basis produced by keeping only one generator from each pair of the full basis and generates the isometry group of a \(1{+}2\) or \(2{+}1\) Minkowski spacetime for which it is also a complete basis. %
Every set shares one and only one unique generator with any other set. 

\(1{+}2\) and \(2{+}1\) Minkowski spacetimes have unique vacuum states where positive and negative frequencies are defined with respect to the asymptotically timelike isometry generator %
and isometries %
preserve the zero frequency~\cite{Carroll19}. %
Consider taking a direct sum of all four of the \(1{+}2\) and \(2{+}1\) projected spaces from our selected signature of the $2{+}2$ embedding space. Direct sums preserve completeness and separability. As a result, such a direct sum produces a space spanned by a new complete basis of isometry generators consisting of the even linear combinations of the separable generators of the projected spaces: \(K^{01} + K^{23}\), \(K^{03} + K^{12}\), and \(J^{02} + J^{13}\). These are the generators of one of the commuting \(\text{SL}(2, \mathbb R)\) groups discussed previously. (The other one can be similarly produced from the same four $1{+}2$ and $2{+}1$ projected spaces with appropriately signed generators.) This direct sum necessarily generates the \(2{+}2\) embedding spacetime on \emph{both} sides of the lightcone (as in Fig.~\ref{fig:embeddingembeddedspace}) since the projections do not favor either side. %

While a double cover of the embedding spacetime with $\mathcal P$- and $\mathcal T$-symmetry does not possess a Cauchy surface, the double cover of the embedding spacetime with broken $\mathcal P$- and $\mathcal T$-symmetry, such as in the direct sum of four three-dimensional Minkowski spacetimes provided, \emph{does} possess a Cauchy surface -- each projected Minkowski spacetime contains an infinite set of Cauchy surfaces in its direct sum decomposition corresponding to its spacelike slices. This direct sum construction produces two non-overlapping copies of the AdS$_3$ embedded spacetimes without use of the $\mathcal P$- and $\mathcal T$-preserving hyperbolic restriction.

As an aside, we note that this construction manifests many of the properties from the work it is inspired from~\cite{Kovalsky23}, which showed that the full $2{+}2$ embedding spacetime acts as a four-dimensional representation of the $\text{SL}(2,\mathbb R)$ group, whereas the embedded AdS$_3$ spacetimes act as the infinite set of three-dimensional representations obtained by projecting down the four-dimensional representation. (States/stress-energy tensors act as the corresponding Lie algebra $\mathfrak{sl}(2,\mathbb R)$, and together these associations produce the complex octonion algebra -- see Appendix~\ref{app:octonionalgebra} for more details.)

  The completeness of the isometry generators in each projected Minkowski space allows any state in the \(2{+}2\) embedding to be specified in terms of the tensor product of its projections. For instance, parametrizing the \(1{+}2\) spaces by \((x,t,x')\) and \((x,t',x')\), and the \(2{+}1\) spaces by \((t,x,t')\) and \((t,x',t')\), means that the Klein-Gordon Fourier modes for the \(2{+}2\) embedding of AdS$_3$ are given by
  \begin{align}
    \label{eq:KGFourierModePhi}
    &\Phi(t,x,t',x') \equiv \\
    &\quad \phi_{+}(t,x,x') \otimes \phi_+(t',x,x') \otimes \phi_-(x,t,t') \otimes \phi_-(x',t,t'), \nonumber
  \end{align}
  and
  \begin{align}
    \label{eq:KGFourierModePi}
    &\Pi(t,x,t',x') \equiv \\
    &\quad \pi_+(t,x,x') \otimes \pi_+(t',x,x') \otimes \pi_-(x,t,t') \otimes \pi_-(x',t,t'), \nonumber
  \end{align}
  where the spaces are tensored together in the order that they were defined and
  \begin{equation}
    \phi_\pm(\tau,\xi_1, \xi_2) = \int \frac{\text d^2 p}{(2 \pi)^2} \frac{1}{\sqrt{2 \omega_{\bs p}}} \left(a_{\bs p} e^{\pm i \bs p \cdot \bs \xi} + a^\dagger_{\bs p} e^{\mp i \bs p \cdot \bs \xi} \right),
  \end{equation}
  \begin{equation}
    \pi_\pm(\tau,\xi_1, \xi_2) = \int \frac{\text d^2 p}{(2 \pi)^2} (\mp i)\sqrt{\frac{\omega_{\bs p}}{2}} \left(a_{\bs p} e^{\pm i \bs p \cdot \bs \xi} - a^\dagger_{\bs p} e^{\mp i \bs p \cdot \bs \xi} \right),
  \end{equation}
  for \(\omega_{\bs p}\) associated with the timelike coordinate \(\tau\) and \(\bs \xi \equiv (\xi_1, \xi_2)\). The isometries of the embedding $2{+}2$ spacetime will preserve the positive and negative frequency modes of the tensor product of the four corresponding Minkowski vacuum states.

Note that, in general, flat spacetimes with multiple timelike isometry generators do not possess a well-defined vacuum state since there is no reason to define frequencies with respect to any particular linear combination of the timelike generators. %
 However, here we have argued that this is not the case for a particular four-dimensional flat spacetime with two timelike generators corresponding to our $2{+}2$ embedding space. This spacetime is distinguished by its lack of \(\mathcal P\) and \(\mathcal T\) symmetry. %

 Since the vacuum state we have defined is for an infinite set ($R<0$) of AdS$_3$ spacetimes, this suggests that its excitations can have preferred radii of curvature $R$ and, therefore, non-stationary states can have dynamic $R$ expectation values. In~\cite{Kovalsky23} it was shown that such evolution, where the $R=0$ case is included, produces a two-dimensional renormalization group flow, and thus forward evolution between the critical points corresponding to zero and non-zero $R$ is irreversible~\cite{Zamolodchikov86}. This suggests the existence of a preferred direction of evolution to this spacetime without \(\mathcal P\) and \(\mathcal T\) symmetry similar to a cosmological arrow of time~\cite{Maxwell1871,Szilard29,Hawking85}.
 
 The decomposition of the embedding $2{+}2$ spacetime into tensor products of projected Minkowski spaces allows for the effects of the curved AdS$_3$ spacetimes with $R<0$ to be independently determined from their properties on these projected flat spacetimes. In particular, examining the projections of the AdS$_3$ embeddings along their hyperbolic restrictions reveals that they correspond to curved trajectories in Minkowski spaces with acceleration proportional to \(l^{-2}\). As a result, the Unruh effect implies that states corresponding to vanishing $R$ are pure while those with finite $R$ are mixed states with temperature proportional to $R$~\cite{Fulling73,Davies75,Unruh76}.

Furthermore, the BTZ black hole spacetime is a discrete quotient of $\text{SL}(2,\mathbb R)$~\cite{Banados93,Maldacena01,Maldacena01_2} such that it covers $\text{SL}(2,\mathbb R)$ if closed timelike curves are permitted from timelike cylindrical coordinates at $r^2<0$~\cite{Nippanikar21}. In this manner, AdS$_3$ without $\mathcal P$ and $\mathcal T$ is locally isometric to such a BTZ black hole, which also has a temperature proportional to \(l^{-2}\)~\cite{Carlip95_2}. As a result, the state we have derived can be equivalently viewed as a consistent vacuum state for an infinite set of BTZ black holes at all temperatures. From this perspective, the preferred direction of evolution agrees with the evaporative evolution of BTZ black holes and captures the eternally radiating property of large thermodynamically stable BTZ black holes. %

The scale invariance in this formulation is responsible for most of its significant results. It can be equivalently described through the existence of the so-called holographic principle in this formulation: any lower  dimensional AdS$_3$ manifold fully determines the properties of its larger dimensional embedding spacetime. It is also in this manner that the emergent property of the original algebraic formulation persists in this quantum field theory; the vacuum state is defined through a ``macroscopic'' embedding space, which is an uniform ensemble of ``microscopic'' embedded spacetimes. Indeed, it is due to these holographic and emergent properties that the isometry generators can produce the embedded spacetimes through their integral curves, and thus make the consistent formulation presented here possible. Notably, this also allows this formulation to be background-independent and for the radius of curvature $R$ of the AdS$_3$ spacetime to become quantum state-dependent.

Emergence and holography have long been considered to be important properties for producing a consistent quantum gravity theory. Their central role in the quantum field theory presented here supports this expectation and suggests that successfuly extending this theory to other non-globally hyperbolic spacetimes will likely require their preservation.

\noindent ---

L{.}~K{.}~Kovalsky~thanks M.~Sarovar, A. Dhumuntarao, and E. Knill for their helpful correspondence during this study. %

This material is based upon work supported by the U.S. Department of Energy, Office of Science, Office of Advanced Scientific Computing Research, under the Accelerated Research in Quantum Computing (ARQC) and Quantum Testbed Pathfinder programs.

Sandia National Laboratories is a multimission laboratory managed and operated by National Technology \& Engineering Solutions of Sandia, LLC, a wholly owned subsidiary of Honeywell International Inc., for the U.S. Department of Energy’s National Nuclear Security Administration under contract DE-NA0003525. 
This paper describes objective technical results and analysis. 
Any subjective views or opinions that might be expressed in the paper do not necessarily represent the views of the U.S. Department of Energy or the United States Government.

\bibliography{biblio}{}
\bibliographystyle{unsrt}

 \clearpage
 \appendix

 \section{Homogeneous Symmetric AdS$_3$'s Killing Vectors}
 \label{app:AdS3Killingvecs}

 The generators of isometries are called Killing vectors. The Killing vector fields on a Riemannian manifold \(M\) form a Lie subalgebra \(\mathfrak g\) of all vector fields on \(M\). For any point \(p \in M\), it can be decomposed through the Cartan decomposition~\cite{Helgason01} into:
 \begin{equation}
   \mathfrak h = \{X \in \mathfrak g: X(p) = 0\}
 \end{equation}
 and
 \begin{equation}
   \mathfrak m = \{X \in \mathfrak g: \nabla X(p) = 0\}.
 \end{equation}

 If \(M\) is a homogeneous space then \(\mathfrak g = \mathfrak h \oplus m\) if and only if \(M\) is a symmetric space~\cite{Olmos14}.

 A family of submanifolds \(N_i \subseteq M\) can be associated with this Lie subalgebra \(\mathfrak g\) by picking a point in \(M\) and setting the tangent space \(T_p N_i\) to be spanned by the Killing vectors. For a symmetric space, \(T_p N_i \cong \mathfrak m\). Such a construction can be used to produce a homogeneous symmetric embedded AdS$_{n+1}$ spacetime in an embedding \(2+n\) Minkowski spacetime for all integer \(n > 1\). For general \(n\), the Killing vector fields produce an overcomplete basis of \(T_p N_i\) since the number of Killing fields is greater than the dimension of the tangent space. \(\mathfrak h\) corresponds to the subset of Killing vectors fields that form an orthogonal space with their degenerate linear combinations. In general, \(\mathfrak h \neq Id\) for some point \(p \in N_i\), where \(Id\) is the identity map, because the overcomplete basis produced by the Killing vector fields of the embedding always have support in the orthogonal space at some point.

 However, for \(n=2\), expressing the embedding \(2{+}2\) Minkowski spacetime as a homogeneous symmetric spacetime reveals only three Killing vectors, equal to the dimension of the embedded AdS$_3$ manifold, which is also a homogeneous symmetric space. Hence, from the point of view of the \(2{+}2\) embedding spacetime, \(\mathfrak h = Id\), and so \(\mathfrak g = Id \oplus T_p N_i\). This means that the Killing vectors form a complete basis, which can generate the AdS$_3$ submanifolds through their integral curves, and these submanifolds must foliate the $2{+}2$ embedding space. The hyperboloid \(-t^2 + x^2 - t'^2 + x'^2 = -l^2\) exactly coincides with these manifolds, where the value of \(l^2{>}0\) determines the particular one.

 This reduction for $n=2$ is particularly useful when combined with the property that projections of homogeneous symmetric spaces are also homogeneous if their kernel is homogeneous. Together, these properties imply that the Killing vectors of an unbiased direct sum of such projections are also unbiased projections of all the Killing vectors from the larger-dimensional homogeneous space. This allows for the four-dimensional homogeneous symmetric spacetime to be faithfully expressed as a direct sum of three-dimensional homogeneous symmetric projected spacetimes. (Considering direct sums of spacetimes with fewer dimensions than three is not useful for our purposes since these cease to have an unique timelike Killing vector.)%

 \section{Relationship to the Octonion Algebra}
 \label{app:octonionalgebra}
 
 The Lie algebra \(\mathfrak g\) from the prior section evaluated at a point \(p\) of the embedding spacetime can be identified with the Weyl algebra isomorphic with the even part of the complexified Clifford algebra \(\text{Cl}^{[0]}_{3,0}(\mathbb C)\), consisting of the generators \(\xi_i\) for \(i \in \{1,2,3\}\) and identity \(\xi_0\), as in~\cite{Kovalsky23}. There, it is shown that the Clifford algebra representation of an element of \(\mathfrak g\), \(T \in \text{Cl}^{[0]}_{3,0}(\mathbb C)\), must be equal to \(T_0 \xi_0 + T_1 \xi_2 \xi_3 + T_2 \xi_1 \xi_3 + T_3 \xi_1 \xi_2\), for \(T_0 \ne 0\) and \(T_i\ne 0\) for some \(i\in\{1,2,3\}\), to be a valid stress-energy tensor for AdS$_3$ in a two-dimensional conformal field theory (\(T_0 \xi_0\) corresponds to \(Id\) and the rest to \(T_p M\) in the Lie algebra generated by the Killing vectors). The development of this two-dimensional conformal field theory requires an extension to the larger \(\text{Cl}_{3,0}(\mathbb C)\) algebra, accomplished through a morphism parametrized by a dimensionless constant, \(\hbar\). Since the stress-energy tensor is proportional to the spacetime's metric \(g_{\mu \nu}\) for isotropic and homogeneous spacetimes (or, equivalently, the curvature tensor is proportional to the metric \(g_{\mu \nu}\), which also follows from the Cartan-Ambrose-Hicks theorem~\cite{Cheeger75}), it follows that the ``coordinates'' \(p = (T_1,T_2,T_3)\) parametrize the scalar curvature \(R\) corresponding to the metric \(g_{\mu \nu}\) in terms of the Killing vectors, \(\xi_i \xi_j\), which generate it through their integral curves, where the relative magnitude of \(p\) compared to \(T_0\) determines the \(l^2\) index of the corresponding AdS$_3$ embedded spacetime. From the BTZ perspective, this means that \(p = (T_1,T_2,T_3)\) parametrize the temperature of the spacetime.

Equivalent to the two-dimensional conformal theory treatment, this algebra can be ``quantized'' to produce a morphism to the Pauli algebra (whose complexification is isomorphic to \(\text{Cl}_{3,0}(\mathbb C)\)) by identifying generator quadratic products with the anticommutator proportional to \(\hbar\). Upon quantization, \(\xi_i \rightarrow \hat \xi_i\), and becomes proportional to a Pauli operator. Hence, the quantized stress-energy tensor, \(\hat T\), corresponds to all unnormalized quantum density operators minus the maximally mixed state. If compactified to the Bloch sphere, the ``pure'' states on the surface of its Bloch sphere correspond to asymptotically flat AdS$_3$ (i.e.~$R \rightarrow 0$; not to be confused with the asymptote of an AdS$_3$ with $R<0$). Otherwise, the metric of AdS$_3$ corresponds to a mixed quantum state.

Expressing \(T \in \text{Cl}^{[0]}_{3,0}(\mathbb C)\) in terms of the non-canonically isomorphic algebra \(\text{Cl}_{4,0}^{[0]}(\mathbb R)\), scales \(T \in \text{Cl}^{[0]}_{4,0}(\mathbb R)\) by an overall factor of \(\hbar/G\) since \(\hat \xi_i \hat \xi_j = \hbar \hat \xi_\mu \hat \xi_\nu / G\)~\cite{Kovalsky23}, but where \(\hbar\) and \(G\) are equal to the canonical quantization constants of the two algebras and so are dimensionless. In AdS$_3$, the stress-energy tensor \(T^{\mu \nu} \propto -R g^{\mu \nu} / G\). This identifies the dimensionless minus scalar curvature, \(-R\), of AdS$_3$ with this dimensionless quantization constant \(\hbar\) (as well as \(G\) with the dimensionless \(G\)). As a result, the \(\text{Cl}_{4,0}^{[0]}(\mathbb R)\) algebra has the characteristics of a ``dual'' in the sense that it is a non-canonically isomorphic algebra to the universal cover \(\text{Cl}^{[0]}_{3,0}(\mathbb C)\) of the ``conformal'' algebra where their canonical quantization constants, \(G\) and \(\hbar\), respectively, are dimensionless analogs of Planck's constant and Einstein's constant. The direct sum \(\text{Cl}_{3,0}^{[0]}(\mathbb C) \oplus \text{Cl}_{4,0}^{[0]}(\mathbb R)\) is isomorphic to the complexified octonion algebra and satisfies many of the dualities of the AdS/CFT correspondence.
 
 Much like \(R < 0\) corresponds to an equivalence class of AdS$_3$ spacetimes and is frequently set to a constant, \(|\hbar| > 0\) corresponds to an equivalence class of qubit Hilbert spaces and is frequently set to a constant. In~\cite{Kovalsky23}, it is shown that this scaling freedom can also be related to the redundant scaling degree of freedom in representations of the central charge in the conformal anomaly. After arbitrarily choosing \(\hbar\) to be non-negative, it is also shown that \(\hbar=0\) and \(\hbar > 0\) correspond to two different critical values in a parametrized irreversible two-dimensional renormalization group flow.

\end{document}